\documentclass[prl,showpacs,twocolumn,superscriptaddress,amsmath,amssymb]{revtex4}

\usepackage[dvips]{epsfig}
\usepackage{color}

\usepackage{dcolumn,amsmath}
 
\begin{document}

\title{Anisotropy induced Feshbach resonances in a quantum dipolar gas of magnetic atoms}

\author{Alexander Petrov}
\affiliation{Department of Physics, Temple University, Philadelphia, Pennsylvania 19122 and National Institute of Standards and Technology, Gaithersburg, Maryland 20899, USA}
\altaffiliation{Alternative address: St. Petersburg Nuclear Physics Institute, Gatchina, 188300; Department of Physics, St.Petersburg State University, 198904, Russia}
\author{Eite Tiesinga}
\affiliation{Joint Quantum Institute,  National Institute of Standards and Technology and
University of Maryland, Gaithersburg, Maryland 20899, USA}
\author{Svetlana Kotochigova}
\email[Corresponding author: ]{skotoch@temple.edu}
\affiliation{Department of Physics, Temple University, Philadelphia, Pennsylvania 19122 and National Institute of Standards and Technology, Gaithersburg, Maryland 20899, USA}

\begin{abstract} 
We  explore the anisotropic nature of Feshbach resonances in the
collision between ultracold magnetic submerged-shell dysprosium atoms,
which can only occur due to couplings to rotating bound states.  This is
in contrast to well-studied alkali-metal atom collisions, where most
Feshbach resonances are hyperfine induced and due to rotation-less
bound states.  Our novel first-principle coupled-channel calculation of
the collisions between open-4f-shell spin-polarized bosonic dysprosium
reveals a striking correlation between the anisotropy due to magnetic
dipole-dipole and electrostatic interactions and the Feshbach spectrum
as a function of an external magnetic field. Over a 20 mT magnetic field
range we predict about a dozen Feshbach resonances and show that the
resonance locations are exquisitely sensitive to the dysprosium isotope.
\end{abstract}

\pacs{03.65.Nk, 31.10.+z, 34.50.-s}

\maketitle

A strongly interacting quantum gas of magnetic atoms, placed in
an optical lattice, provides the opportunity to examine strongly
correlated matter, creating a platform to explore exotic many-body
phases known in solids, quantum ferrofluids, quantum liquid crystals,
and supersolids \cite{Fregoso2009,BLev2009}.  Recent experimental advances
\cite{Pfau2006,McClelland,Weinstein2009,Doyle2010,BLev2010,BLev2011,Sukachev,BLev2012}
in trapping and cooling magnetic atoms pave the way towards these goals.

In general, interactions between magnetic atoms are orientationally
dependent or anisotropic.  At room temperature anisotropic interactions
are much smaller than kinetic energies and other major interactions
between atoms, therefore can be ignored. The situation is different
for an ultracold gas of atoms with a large magnetic moment. It was, for
example, demonstrated that the anisotropy due to magnetic  dipole-dipole
interactions between ultracold chromium atoms leads to an anisotropic
deformation of a Bose Einstein condensate (BEC) \cite{Pfau2005}. Moreover,
anisotropy plays a dominant role in collisional
relaxation of ultracold atoms with large magnetic moments
\cite{Krems,Weinstein2009,Doyle2010,Hensler2003,Hancox2004,BLev2010,Kotochigova2011}.

\begin{figure}[b]
\includegraphics[scale=0.3,trim=0 28 0 53,clip]{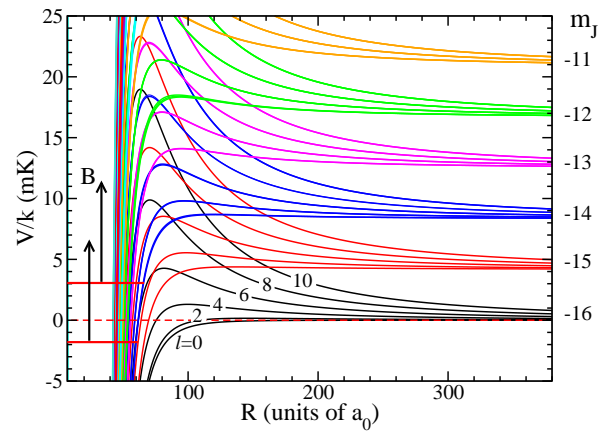}
\caption{(Color online) Potential energy curves for a
$^{164}$Dy+$^{164}$Dy collision in an external magnetic field $B$ as a
function of internuclear separation.  The (red) dashed line with zero
energy indicates the energy of the entrance channel.  Two Feshbach
resonances are schematically shown by (red) horizontal lines, which end at the classical
outer turning point of a closed channel.  
Their energy increases, indicated by arrows, with magnetic field and a
resonance occurs when this energy equals the entrance-channel energy.
There are 91 diagonal potential matrix elements
for channels  $|(j_1j_2)jm_j,\ell m_\ell\rangle$ with $m_j+m_\ell=-16$ and even
$\ell\leq10$. We use $B=50$ G.  The curves are colored by their $m_j$
value, while for $m_j=-16$ curves their $\ell$ value is indicated.
Here 1 G=0.1 mT,
$a_0=0.0529177$ nm is the Bohr radius, and $k=1.38065\cdot 10^{-23}$
J/K is the Boltzmann constant.
}
\label{channels}
\end{figure}

In this Letter we pursue ideas for using anisotropic magnetic and
dispersion interactions to control collisions of ultracold
magnetic atoms by using Feshbach resonances \cite{Tiesinga}.
Resonances, shown schematically in Fig.~\ref{channels}, appear when
the energy of ``embedded'' bound states cross the energy of the entrance channel
or initial scattering state.  The embedded state is a level of a
potential dissociating to a closed channel whose asymptotic energy is
larger than that of the entrance channel. Coupling with the entrance
channel leads to a resonance.

Feshbach resonances make it possible to convert a weakly interacting
gas of atoms into one that is strongly interacting and along the way
promise to make available many of the collective many-body states mentioned
above. Alternatively, interactions can be turned off all together to
create an ideal Fermi or Bose gas, for which thermodynamic properties are known
analytically. Feshbach resonances can also be used to create BECs of
weakly-bound molecules \cite{Kohler2006}, which can be optically stabilized
to deeply-bound molecules \cite{Ni2008}. For fermionic
atoms the BCS-BEC phase transition \cite{BCS} and universal many-body
behavior of strongly interacting magnetic atoms can be studied via Feshbach resonances.
Finally, three-body Efimov physics \cite{Kraemer2006} can be explored.

The most promising atoms to look for the effect of anisotropy
on collisions are submerged-shell atoms, which have an electronic
configuration with an unfilled inner shell shielded by a closed outer
shell. In particular, we are interested in the rare-earth dysprosium
(Dy) atom with a $^5{\rm I}_8$ ground-state, a total angular momentum
$j=8$, and a large magnetic moment of $\approx 10\mu_B$, for which the
electron spins of the inner 4f$^{10}$ shell are aligned such that its
orbital angular momentum is maximal and largely unquenched.  Here $\mu_B$ is the
Bohr magneton.  As a result, Dy's magnetic and electrostatic properties
are highly anisotropic.  A quantitative description of the collision
between two dysprosium atoms is challenging.  For example, our previous
study \cite{Kotochigova2011} showed that there are 153 Born-Oppenheimer
potentials that dissociate to the ground $^5{\rm I}_8+{}^5{\rm I}_8$
state.

We present a first-principle coupled-channel model allowing us
to calculate anisotropy-induced magnetic Feshbach-resonance spectra of
bosonic Dy atoms.  The model treats the Zeeman, magnetic dipole-dipole,
and electrostatic isotropic and anisotropic dispersion interactions
equally.  Bosonic Dy isotopes have zero nuclear spin.
Thus, there is no nuclear hyperfine structure and only Zeeman
splittings remain. The weak quadrupole-quadrupole interaction 
\cite{Kotochigova2011} is included for completeness.

We already note that Feshbach resonances in rare-earth
magnetic atoms are different in nature than those in alkali-metal atom
collisions.  As pointed out in Ref.~\cite{Kotochigova2011} for a
coupled-channel calculation there exists at most one channel with zero
relative nuclear orbital angular momentum $\vec \ell$, which for ultra-cold collisions
is also the entrance channel. Consequently, resonances 
occur due to anisotropic coupling to bound states with non-zero $\ell$.

The focus of this Letter is on ultra-cold collisions of atoms prepared
in the energetically-lowest Zeeman state $j=8$ and projection $m=-8$.
Inelastic exothermic atom-atom processes, where the spin projection of one or both
of the atoms changes, are absent and, consequently, Feshbach resonances
can be readily observed.  

We start by setting up the Hamiltonian, interatomic potentials, and
channel basis for two bosonic $^5{\rm I}_8$ Dy atoms with
zero nuclear spin.  This Hamiltonian assuming a magnetic field $B$ along the 
$\hat z$ direction is
\begin{equation}
H = -\frac{\hbar^2}{2\mu_r}\frac{d^2}{dR^2} + 
      \frac{\vec \ell^2}{2\mu_rR^2} + H_Z +V(\vec R,\tau)  \,,
\label{ham}
\end{equation}
where $\vec R$ describes the orientation of and separation between
the two atoms. The first two terms are the radial kinetic and
rotational energy operators, respectively. The Zeeman interaction is
$H_Z=g_s\mu_B (j_{1z}+j_{2z}) B$ with $g_s=1.24159$ the g-factor of
Dy \cite{NIST} and $j_{iz}$ is the $z$ component of the angular momentum
operator $\vec\jmath_i$ of atom $i=1,2$.  The electronic Hamiltonian,
including nuclear repulsion, $V(\vec R,\tau)$ is anisotropic and $\tau$
labels the electronic variables. Finally, $\mu_r$ is the reduced mass
and for $R\to\infty$ the interaction $V(\vec R,\tau)\to0$.

Our coupled-channels calculations \cite{cc} are performed in the
atomic basis $|(j_1j_2)jm_j,\ell m_\ell\rangle\equiv Y_{\ell
m_\ell}(\theta,\phi)|(j_1j_2)jm_j\rangle$, where $Y_{\ell m_\ell}(\theta,\phi)$
is a spherical harmonic and angles $\theta$ and $\phi$ give the
orientation of the internuclear axis relative to the magnetic field
direction. In this basis the Zeeman and rotational interaction are
diagonal with energies $g_s\mu_Bm_jB+\hbar^2\ell(\ell+1)/(2\mu_rR^2)$.
Coupling between the basis states is due to $V(\vec R,\tau)$ and will
be discussed in detail below.  Excited atomic states, for example
those with $j_i\ne 8$, are not included as their internal energy is
sufficiently high that the effects of coupling to these states is
negligible.  The Hamiltonian $H$ conserves $M_{\rm tot}=m_j+m_\ell$
and is invariant under the parity operation so that only even (odd)
$\ell$ are coupled. For homonuclear collisions only basis states with
even $j+\ell$ exist. Figure \ref{channels} shows an example of the
long-range diagonal matrix elements in the atomic basis of the sum
of the rotational, Zeeman, and electronic Hamiltonian.  We have used
$M_{\rm tot}=-16$ and even $\ell\leq 10$.  In fact, only the potentials
dissociating to the six energetically-lowest Zeeman states are shown. The
large number of potentials indicates the large number of resonances that,
in principle, are possible.

Coupling between basis states is due to $V(\vec R,\tau)$.  It is convenient
to first evaluate this operator in a molecular basis with body-fixed
projection quantum numbers defined with respect to the internuclear axis.
We use the molecular basis $|(j_1j_2)j\Omega\rangle$ with projection
$\Omega$ of $\vec \jmath$ along the internuclear axis.  For Dy$_2$ the
matrix elements of $V(\vec R,\tau)$ conserve the projection $\Omega$
but not the length $j$.  The eigenenergies of $V(\vec R,\tau)$ at
each value of $R$ are the adiabatic (relativistic) Born-Oppenheimer
potentials \cite{Herzberg,Lefebre1986}. Typically, these potentials
$U_{n|\Omega|\sigma}(R)$ are obtained from an 
electronic structure calculation and labeled by $n|\Omega|_{\sigma}^{\pm}$,
where $|\Omega|$ is the absolute value of $\Omega$, $\sigma=g/u$ is
the {\it gerade/ungerade} symmetry of the electronic wavefunction,
and $n=1,2,\cdots$ labels curves of the same $|\Omega|_{g/u}^{\pm}$
in order of increasing energy.  For bosonic Dy$_2$ the 81 {\it gerade}
states are superpositions of even $j$, while the 72 {\it ungerade}
states are superpositions of odd $j$.

For $R> 27 a_0$, beyond the Le Roy radius where the atomic electron
clouds have negligible overlap, we assume that $V(\vec R,\tau)$ is the
sum of the magnetic dipole-dipole, $V_{\mu \mu}(\vec R) \propto 1/R^3$,
the electric quadrupole-quadrupole, $V_{QQ}(\vec R) \propto 1/R^5$,
and the van-der-Waals dispersion $V_{\rm disp}(\vec R) \propto 1/R^6$
interaction.  Reference~\cite{Kotochigova2011} reported the matrix
elements of the operator $V_{\rm disp}(\vec R)$ in the molecular basis
and tabulated the adiabatic $C_{6, n\Omega\sigma}$ dispersion
coefficients obtained by diagonalizing $V_{\rm disp}(\vec R)$. Crucially,
the eigenfunctions of $V_{\rm disp}(\vec R)$ are independent of $R$.

At shorter range coupling between basis states is more complex.
Rather than determining all Born-Oppenheimer potentials,
we have opted for the following approach.
First, we calculate the single {\it gerade} potential $U_{16g}(R)$
with the maximal projection $\Omega=16$ using a coupled-cluster
method with single, double, and perturbative triple excitations
(CCSD(T)) \cite{uccsdt} together with the scalar relativistic
Stuttgart ECP28MWB pseudopotential and associated atomic bases sets
(14s,13p,10d,8f,6g)/[10s8p5d4f3g].  The potential has a minimum
at $R_e= 8.771 a_0$ with depth $D_e/(hc)= 785.7$
cm$^{-1}$. For $^{164}$Dy$_2$ it has a $\omega_e/(hc)= 25.6$ cm$^{-1}$
and contains 71 bound states.  (We omit the $n=1$ label in $U_{16g}(R)$.)
We then assume that the $R< 27 a_0$ electronic wavefunctions of the
Born-Oppenheimer potentials are the same as those determined by the
dispersion interaction, and the relation between energies of the {\it
ab-initio} potentials is the same as for its $C_6$ coefficient. Hence,
the adiabatic potentials satisfy $U_{n\Omega\sigma}(R)/U_{n'\Omega\sigma
}(R) = C_{6,n\Omega\sigma}/C_{6,n'\Omega'\sigma}$  for $R< 27 a_0$ {\it
and} with eigenfunctions as determined by the dispersion interaction.
Equivalenty, this allows us to write $V(\vec
R,\tau)$ in terms of $U_{16g}(R)$ as
\begin{eqnarray}
V(\vec R,\tau ) = V_{\mu\mu}(\vec R)+V_{QQ}(\vec R) 
        +\frac{U_{16g}(R)}{-C_{6,16g}/R^6} V_{\rm disp}(\vec R)
\end{eqnarray}
for any $R$.  This approach reduces the number of 
independent short-range potentials to one.

\begin{figure}
\includegraphics[scale=0.3]{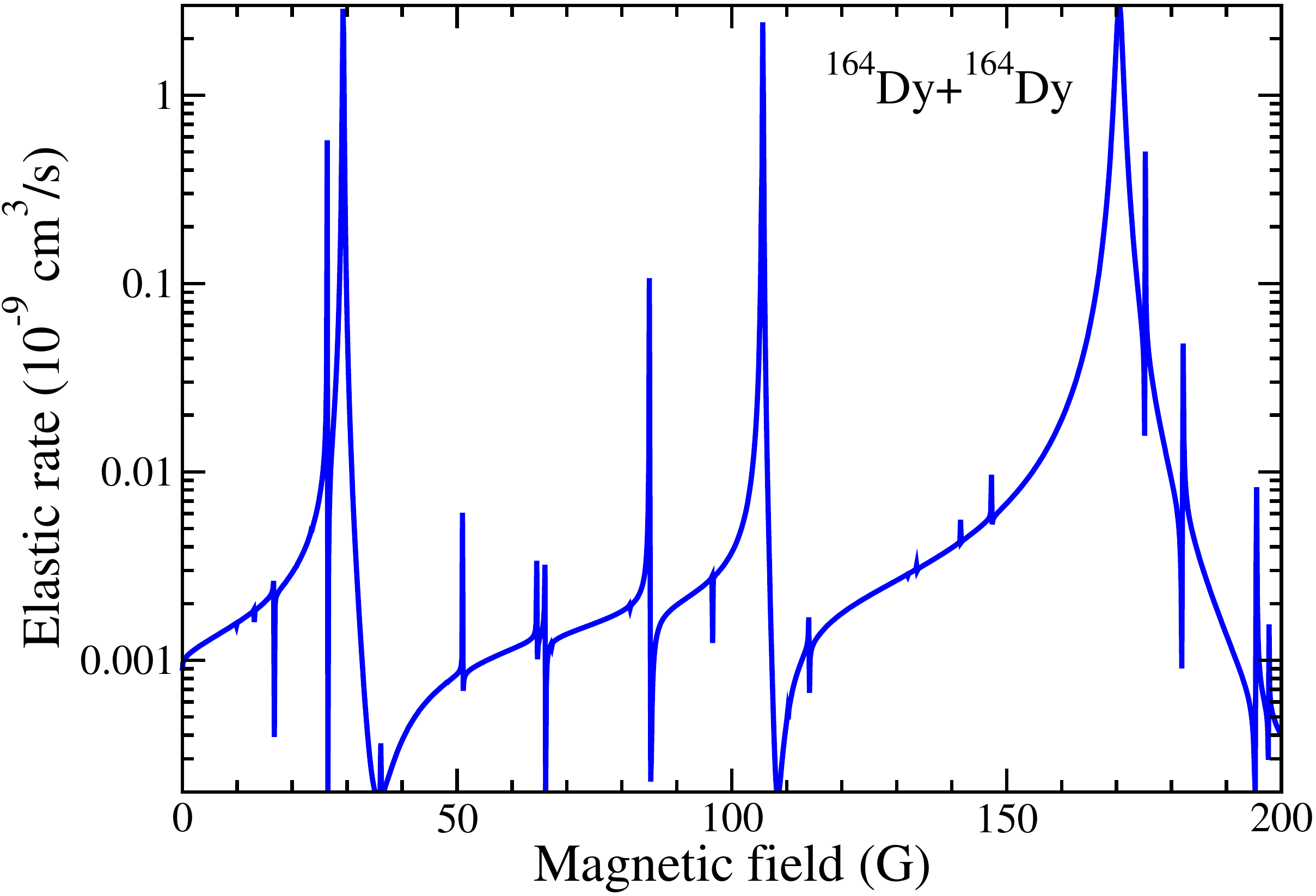}
\caption{Elastic rate coefficient of $m = -8$
$^{164}$Dy collisions as a function of magnetic field using a collision
energy of $E/k=$ 30 nK. Partial waves $\ell$ up to 10 are included.}
\label{feshbach}
\end{figure}

\begin{table}[b]
\caption{Dispersion coefficients $c^{(i)}_{k}$ in units of $E_ha_0^6$, where
$E_h=4.35974 \times 10^{-18}$ J is the Hartree.
The strength of the magnetic dipole-dipole and quadrupole-qudrupole interaction are
$c_{\mu \mu}=-5.0269 \cdot 10^{-3}$ $E_ha_0^3$ and $c_{QQ}=9.5719\cdot 10^{-8}$ $E_ha_0^5$,
respectively.
} 
\begin{tabular}{c|cccc}
\hline
       $k\backslash i$     &  1    & 2   & 3    \\ 
       \hline
0      &$-1873.4$ &  $3.57 \cdot 10^{-3}$ &  $-6.82 \cdot 10^{-6}$   \\
2     & $-0.1680$ &  $5.06 \cdot 10^{-3}$ & $-8.15 \cdot 10^{-6} $     \\
4 &  $-6.56 \cdot 10^{-5}$    
\end{tabular}

\label{tcoef}
\end{table}

For practical reasons it is advantageous to write $V_{\rm disp}(\vec R)$ as a
sum of spherical tensor operators in the laboratory frame.  That is 
\begin{equation}
  V_{\rm disp}(\vec R) = \frac{1}{R^6}
        \sum_{kq} \sum_i c^{(i)}_{k} \, (-1)^q C_{k,-q}(\theta,\phi)
        T_{kq}^{(i)} \,,
\label{tensor} \end{equation} where
$C_{kq}(\theta,\phi)=\sqrt{4\pi/(2k+1)}Y_{kq}(\theta,\phi)$ and the
spherical tensors $T_{kq}^{(i)}$ of rank $k$ and with components $q$
are defined by 
\begin{eqnarray} &&T^{(1)}_{00} = I ,\quad 
T^{(1)}_{2q} = [j_1 \otimes j_1 ]_{2q} + [j_2 \otimes j_2 ]_{2q} \,, \nonumber\\
&&T^{(2)}_{2q} =[j_1 \otimes j_2 ]_{2q}  \,, \quad
                                            T^{(2)}_{00} =[j_1 \otimes j_2 ]_{00} \,,  \nonumber  \\
&&T^{(1)}_{4q} = \left[ [j_1 \otimes j_1 ]_2 \otimes  [j_2 \otimes j_2 ]_2 \right]_{4q}  \,, \label{quart}\\  
&&T^{(3)}_{2q} = \left[ [j_1 \otimes j_1 ]_2 \otimes  [j_2 \otimes j_2 ]_2 \right]_{2q}  \,,\nonumber \\ 
&&T^{(3)}_{00} = \left[ [j_1 \otimes j_1 ]_2 \otimes  [j_2 \otimes j_2 ]_2 \right]_{00} \,, \nonumber
\end{eqnarray} 
where $I$ is the identity operator and $ [j \otimes j']_{kq}$
denotes a tensor product of angular momentum operators $\vec \jmath$
and $\vec \jmath'$ coupled to an operator of rank $k$ and component
$q$ \cite{Greene2003}.  The higher-order tensor operators are constructed in an
analogous manner.  Equation \ref{quart} has three, three, and one tensors
$T^{(i)}_{kq}$ of rank $k=0$, 2, and 4, respectively.  The dispersion
coefficients $c^{(i)}_{k}$, listed in Table \ref{tcoef}, determine the strength of each term.
Their order in Eq.~\ref{quart} is by decreasing absolute value.  The isotropic
$T^{(1)}_{00}$ term is the largest by far with the strongest anisotropic
contribution from a dipolar (rank-2) operator constructed from the angular momentum
of only one atom coupled to the rotation of the molecule.  Finally, the magnetic
dipole-dipole and quadrupole-quadrupole interaction are
\begin{eqnarray}
         V_{\mu\mu}(\vec R) &=& \frac{1}{R^3}c_{\mu\mu} \sum_q (-1)^q
         C_{2,-q}(\theta,\phi) T^{(2)}_{2q} \\ V_{QQ}(\vec R) &=&
         \frac{1}{R^5}c_{QQ} \sum_q (-1)^q C_{4,-q}(\theta,\phi)
         T^{(1)}_{4q} \,,
\end{eqnarray} respectively.  Their strengths are listed in Table
\ref{tcoef}.

Figure \ref{feshbach} shows the elastic rate coefficient for the collision
between two $m=-8$ $^{164}$Dy atoms at a collision energy of $E/k=30$ nK.
We use $M_{\rm tot}=-16$ and include channels with even $\ell$ up to ten
leading to a close-coupling calculation with 91 channels. Fields up to
$B=200$ G fall comfortably within the experimentally
accessible values \cite{Tiesinga}.   
The graph shows a dozen of Feshbach resonances; some are broad, many
are very narrow.  By performing calculations
that include fewer partial waves we have observed that the resonances can
not be labeled by a single partial wave. For example, the three broad
resonances at $B\approx 30$ G, 110 G, and 170 G are already present when
only $\ell=0$, 2, and 4 channels are included.  Their locations, however,
shift significantly when higher $\ell$ channels are included and only
converge to within a few Gauss when $\ell=8$ channels are included.
In general, we find that the magnetic-field location of a resonance that
appears when channels with partial wave $\ell$ are included stabilizes
when channels up to $\ell+4$ are included.  

We stress that this behavior with increasing number of
channels is unlike that observed in alkali-metal atom collisions
\cite{Tiesinga} or even in collisions of strongly magnetic chromium
atoms \cite{Simoni2005}. For these atoms resonances do not shift by
more than a few Gauss when additional partial waves are added. Hence,
resonances can be labeled by a partial wave quantum number. For dysprosium
the anisotropic interactions are so strong that states with different
partial waves are strongly mixed.
We have also studied the effect of the uncertainty in the depth of the
$\Omega=16$ Born-Oppenheimer potential.  The depth $D_e$ was changed by
adding a localized correction to $U_{16g}(R)$ that does not modify its
long-range potential.  A depth change by no more than
10 cm$^{-1}$ changes its number of bound states by one.
Changing the depth by smaller amounts changes the resonance
spectrum non-trivially. For example resonance widths can
be modified dramatically and rate coefficients with broad resonances
that appear when $d$-wave channels are included can be observed.

\begin{figure}
\includegraphics[scale=0.3]{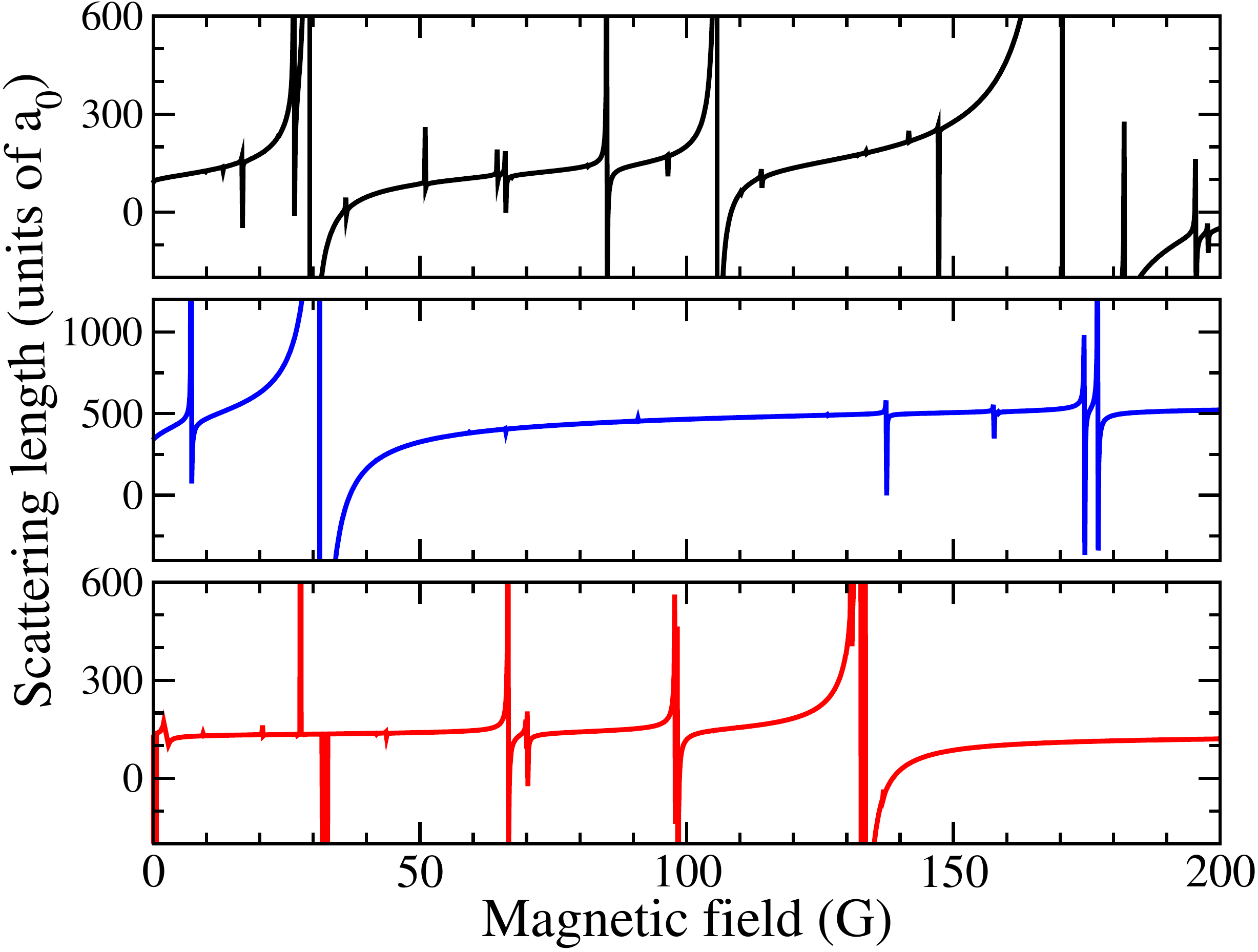}
\caption{Scattering length of $m=-8$ $^{164}$Dy atoms as a
function of magnetic field with and without the magnetic dipole-dipole
or the anisotropic contribution of the dispersion interaction.  The top
panel shows the case when all interactions are included.  At $B=0$
the scattering length is $89 a_0$.  The middle and bottom panels are
obtained when the dispersion and magnetic dipole-dipole anisotropy is set
to zero, respectively. Even waves $\ell$ up to 10 are included.
}
\label{anisotropy}
\end{figure}

The precise form of short-range potential and dispersion coefficients
are not known. A few percent uncertainty is not unrealistic.  For this
Letter we have constructed potentials that lead to a positive $B=0$
scattering length $a$ for $^{164}$Dy atoms in the $m=-8$ state. This
choice is suggested by the recent observation of a Bose condensed gas of
$^{164}$Dy atoms at nearly zero magnetic field.  It thus possesses a
positive scattering length at this field \cite{BLev2011}.  Moreover,
we chose the scattering length to be approximately equal to the mean
scattering length \cite{Gribakin1993} for (fictitious) scattering of a
van der Waals potential with a $C_6$ coefficient equal to the isotropic
dispersion coefficient.

To further elucidate the effect of anisotropy, Fig.~\ref{anisotropy}
shows the scattering length of $m=-8$ $^{164}$Dy collisions as a
function of magnetic field when parts of the anisotropy are turned off.
The top panel displays the case when all interactions are included
corresponding to the elastic scattering described in Fig.~\ref{feshbach}.  
The bottom two panels show
the effect of turning of the anisotropy in the dispersion and magnetic
dipole-dipole interaction, respectively.  The resonance spectra in the
three panels are quite distinct. The number of resonances differs and,
with one exception, the resonances are narrower. 

\begin{figure}
\includegraphics[scale=0.3]{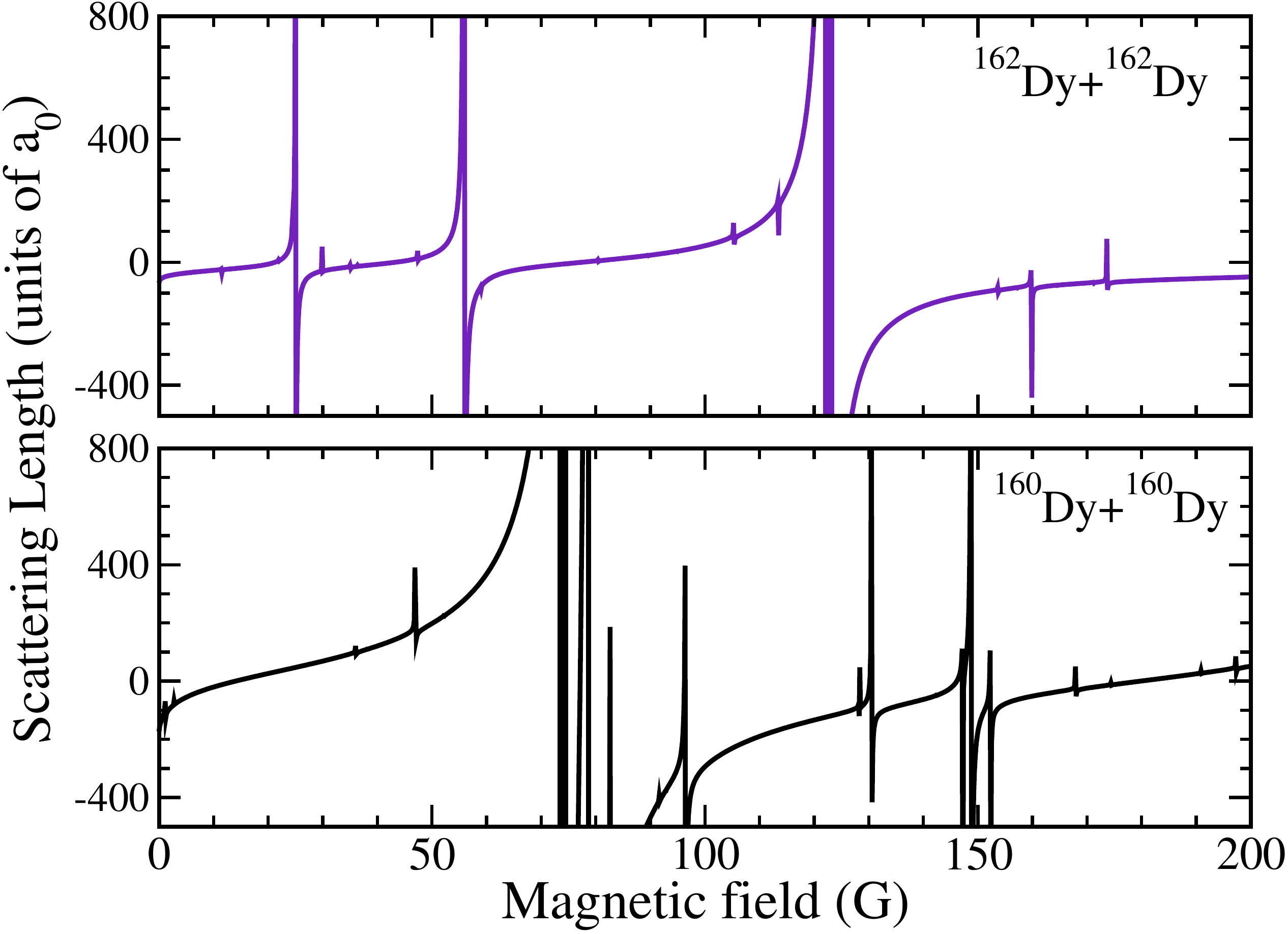}
\caption{Scattering length as a function of magnetic field for the 
bosonic isotopes $^{162}$Dy (top panel) and $^{160}$Dy (bottom panel). The
magnetic state and number of included $\ell$ is as in Fig.~\ref{anisotropy}.}
\label{isotopes}
\end{figure}

Finally, Fig.~\ref{isotopes} illustrates the effect of changing
to different bosonic Dy isotopes.  Since to good approximation
Born-Oppenheimer potentials do not depend on the isotope, we have solved
the coupled-channels equations using the appropriate reduced mass $\mu_r$.
This observation has been used in understanding relationships between
scattering lengths of isotopic combinations of spin-less ytterbium
\cite{Julienne}, while its limitations for Lithium Feshbach resonances
have been studied in Ref.~\cite{Lithium}.  The field dependence of the
scattering length changes from $^{160}$Dy to the $^{162}$Dy isotope.
Measurement of resonance locations in different isotopes will be
invaluable in understanding the scattering of dysprosium.

{\it Conclusion.} Applying a full coupled-channels calculation for
ultracold atom-atom collisions, we have shown that the origin of Feshbach
resonances in interactions between ultracold rare-earth atoms with large
magnetic moments result from strong scattering anisotropies. Consequently,
by tuning an applied magnetic field we predict that it will be possible
to observe resonances and control collisions even for atoms with zero
nuclear spin. This study is the first predictions of a Feshbach resonance
spectrum for  rare-earth atoms.

We have investigated the effects of different short-range and long-range
anisotropic potentials as well as different isotopes on the scattering
length of ultracold Dy atoms as a function of magnetic field strength.
To optimize the potentials we must await experimental observations of
resonances from multiple isotopic combinations.

\section{Acknowledgments} This work is supported by grants of the AFOSR
grant No.~FA 9550-11-1-0243 and NSF No.~PHY-1005453.


\begin{thebibliography}{99}
\bibitem{cc} A close-coupling calculation is a numerical method 
to find scattering solutions of a finite set of coupled radial
Schr\"odinger equations.
\end{thebibliography}
\end{document}